\documentclass[aps,prb,amsmath,amssymb,titlepage,superscriptaddress,showpacs, 12pt]{revtex4-1}
\usepackage[english]{babel}
\usepackage{graphicx}
\usepackage{amsmath}
\usepackage{amssymb}
\usepackage[outdir=./]{epstopdf}
\usepackage{amsfonts}
\usepackage{bm}
\usepackage{mathtools}
\usepackage{times}
\usepackage{hyperref}
\usepackage{color}
\usepackage[dvipsnames]{xcolor}
\usepackage{soul}
\newcommand{\I}{\mathrm{i}}

\newcommand{\bsplit}[1]{\begin{equation} \begin{split} #1 \end{split} \end{equation}}

\newcommand{\astcycl}{\mathrlap{\kern0.085em{\circlearrowright}}\ast}
\newcommand{\taucycl}{\mathrlap{\kern0.42em{\bullet}}\circlearrowright}
\newcommand{\Neel}{N\'{e}el }

\newcommand{\ca}{\tilde c^{\phantom{\dagger}}}
\newcommand{\cc}{\tilde c^\dagger}
\newcommand{\n}{\tilde n}

\long\def\/*#1*/{}

\def\<{\left\langle}
\def\>{\right\rangle}

\newcommand{\vS}{\textbf{S}}

\usepackage{tikz}
\usetikzlibrary{arrows,decorations.pathmorphing}

\renewcommand{\vec}[1]{\mathbf{#1}}
\begin{document}

\title{Ultrafast coupled charge and spin dynamics in strongly correlated NiO}

\date{\today} 

\author{Konrad Gillmeister}
\affiliation{Institute of Physics, Martin-Luther-Universit\"{a}t Halle-Wittenberg, 06120 Halle, Germany}
\author{Denis Gole\v{z}}
\affiliation{Center for Computational Quantum Physics, Flatiron Institute, 162 Fifth Avenue, New York, NY 10010, USA}
\affiliation{Department of Physics, University of Fribourg, 1700 Fribourg, Switzerland}
\author{Cheng-Tien Chiang}
\affiliation {Institute of Physics, Martin-Luther-Universit\"{a}t Halle-Wittenberg, 06120 Halle, Germany}
\author{Nikolaj Bittner}
\affiliation{Department of Physics, University of Fribourg, 1700 Fribourg, Switzerland}
\author{Yaroslav Pavlyukh}
\affiliation {Department of Physics, Technische Universit\"{a}t Kaiserslautern, 67653 Kaiserslautern, Germany}
\author{Jamal Berakdar}
\affiliation {Institute of Physics, Martin-Luther-Universit\"{a}t Halle-Wittenberg, 06120 Halle, Germany}
\author{Philipp Werner}
\affiliation{Department of Physics, University of Fribourg, 1700 Fribourg, Switzerland}
\author{Wolf Widdra}
\affiliation {Institute of Physics, Martin-Luther-Universit\"{a}t Halle-Wittenberg, 06120 Halle, Germany}
\affiliation {Max Planck Institute of Microstructure Physics, 06120 Halle, Germany}

\begin{abstract}
Charge excitations across an electronic band gap play an important role in opto-electronics and light harvesting. 
In contrast to conventional semiconductors, studies of above-band-gap photoexcitations in strongly correlated materials are still in their infancy. Here we reveal the ultrafast dynamics controlled by Hund's physics in strongly correlated photo-excited NiO.
By combining time-resolved two-photon photoemission experiments with state-of-the-art numerical calculations, an ultrafast ($\lesssim$ 10\,fs) relaxation due to Hund excitations and related photo-induced in-gap states  
are identified. Remarkably, the weight of these in-gap states displays long-lived coherent THz oscillations up to 2\,ps at low temperature. The frequency of these oscillations corresponds to the strength of the antiferromagnetic superexchange interaction in NiO and their lifetime vanishes as the N\'eel temperature is approached. 
Numerical simulations of a two-band $t$-$J$ model reveal that the THz oscillations originate from the interplay between  local many-body excitations and long-range antiferromagnetic order.

\end{abstract}
\maketitle

Employing light to understand and control ordered states in solid-state systems has been a challenge at the frontier of modern science and technology.
The optical control of spins in ferromagnets is an especially important research area, related to the ultrafast writing of magnetic information, which has recently been extended to antiferromagnets~(AFM)~\cite{nemec_antiferromagnetic_2018}.
An interesting class of AFM materials is the family of strongly correlated electron systems, where the local electron correlations lead to insulating behavior and simultaneously stabilize the long-range magnetic ordering via Anderson's superexchange mechanism~\cite{anderson_antiferromagnetism_1950}.
The complex electronic structure of correlated materials provides coupling mechanisms between the magnetic order and the orbital, spin and lattice degrees of freedom, which leads to numerous competing metastable phases accessible via ultrafast photoexcitation.
In recent years, several exciting scenarios for the ultrafast manipulation of ordered states have been reported, including  possible high-temperature superconducting states in light-driven cuprates~\cite{fausti2011,kaiser_optical_2014} and
fullerides~\cite{mitrano2016}, hidden states in 1T-TaS$_2$~\cite{stojchevska_ultrafast_2014}, as well as photo-induced non-thermal magnetic and orbital-ordered states~\cite{li2018,ono_double-exchange_2017,atsushi2018}.
Less attention has been paid to the feedback mechanisms between the long-range electronic or magnetic ordering and the dynamics of the photoexcited charge carriers.

In correlated transition metal compounds, the strong electron-electron repulsion splits the $d$ bands into an occupied lower Hubbard band~(LHB) and an empty upper Hubbard band~(UHB). 
Optical excitations at photon energies larger than the Mott gap between the LHB and the UHB leads to photo-doping of the UHB.
Since the Mott gap is the largest energy scale in the system, the recombination of excited charge carriers is suppressed, which results in long-lived metastable states. 
Such a transient electronic structure can be experimentally probed by photoemission spectroscopy.
Using pump-probe time-resolved two-photon photoemission (2PPE) with two independently tunable ultraviolet (UV) sources, we study the fate of the excited electrons in the UHB of the transition-metal oxide~(TMO) NiO triggered by photoexcitation across the band gap.
Upon above-band-gap optical excitation, we find: a) an ultrafast decay of the excited electrons within less than 10\,fs, and b) a photo-induced in-gap state well below the bottom of the UHB of the originally undoped system.
These two effects represent the non-equilibrium manifestation of the Hund's coupling in NiO.
We demonstrate that the many-body in-gap state is coupled to the antiferromagnetic spin background providing conditions for the coherent THz oscillations of the 2PPE signal.
The frequency of the THz oscillations corresponds to the superexchange interaction in NiO, and its amplitude decreases to zero with a square-root dependence when approaching the \Neel temperature ($T_{N}$).
These experimental findings are complemented by theoretical modeling based on a two-band $t$-$J$ model, which reveals the strong coupling between the excited Hund states and the antiferromagnetic background as the origin of the coherent oscillations.

One of the first systematic classifications of TMOs is due to Zaanen, Sawatzky, and Allen (ZSA)~\cite{zaanen_band_1985}. They suggested that the electronic properties of TMOs are essentially determined by two parameters: the Hubbard $U$ and the charge transfer $\Delta$. Mott-Hubbard insulators are characterized by $U < \Delta$.
In these systems, the oxygen (ligand $L$) $p$-band is located well below the TM $d$ band and plays a minor role in the low-energy dynamics. In charge-transfer insulators, $\Delta$ is smaller than $U$, and the $p$-band is located between the LHB and the UHB, as schematically depicted in Fig.~1(a).
  
  \begin{figure}
  \includegraphics[width=0.9\linewidth]{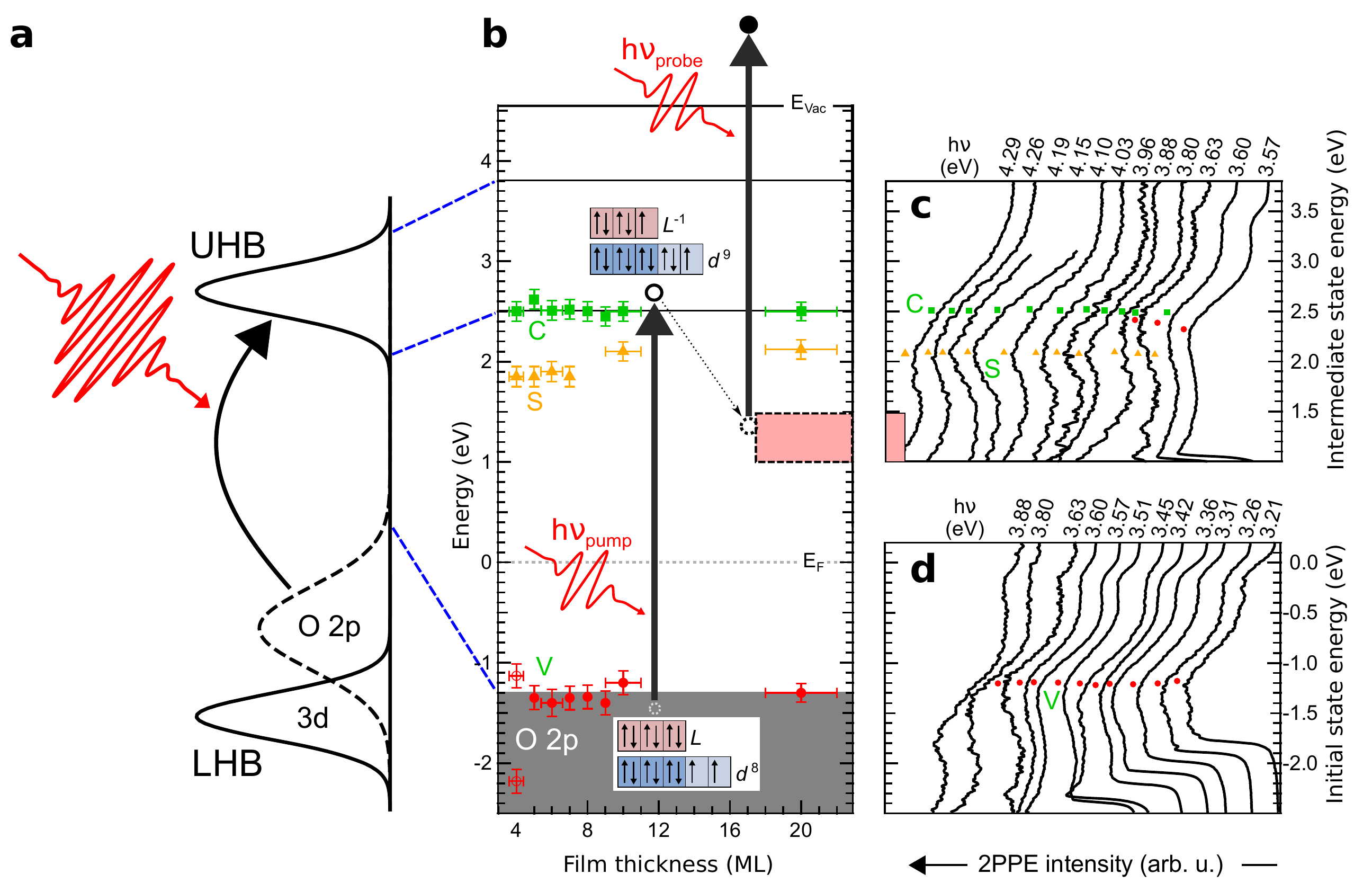}
 \caption{\textbf{a}, Schematics of photoexcitation in an ideal charge transfer
   insulator. LHB (UHB): lower (upper) Hubbard bands formed by the 3$d$
   electrons; O 2$p$: oxygen valence bands. 
   \textbf{b}, Electronic structure of ultrathin NiO films on Ag(001) with a
   thickness from 4 to 20\,ML. Black arrows illustrate the pump ($h\nu_\text{pump}$)
   and probe ($h\nu_\text{probe}$) processes in time-resolved 2PPE. The relevant
   electronic states are: $d^{8}L$ ground states of the LHB (V, bottom),
   photoexcited $d^{9}L^{-1}$ states of the UHB (S and C, top), as well as 
   photo-induced in-gap states associated with the THz oscillations (red filled).
   Red boxes of $L$
   ($L^{-1}$) in the insets indicate the ligand 2$p$ states without (with) a photo-hole, whereas dark
   (light) blue boxes represent the $t_{2g}$ ($e_{g}$) orbitals of the 3$d$
   states. 
   \textbf{c} and \textbf{d}, one-color 2PPE spectra of 10\,ML NiO films with a
   variable photon energy $h\nu$.}
\label{2PPE}
\end{figure}

In reality, the band structure of NiO is much more convoluted and intricate,
resulting from the $d^8$ ground state electronic configuration with formally
fully occupied $t_{2g}$ and half-filled $e_g$ orbitals as indicated in the dark
and light blue insets of Fig.~\ref{2PPE}(b), respectively. The established energy scales for
NiO are~\cite{sawatzky_magnitude_1984,fujimori_multielectron_1984}:
\begin{eqnarray*}
\Delta&=&E(d^9L^{-1})-E(d^8)\sim4~\text{eV},\\
U&=&E(d^7)+E(d^9)-2E(d^8)\sim7.5~\text{eV},
\end{eqnarray*} 
where $L^{-1}$ denotes a hole at the ligand site.
Accordingly, NiO is an intermediate charge-transfer case with the $p$-band located in the energy range of the LHB.
The latter situation leads to a strong hybridization between the $2p$ and $3d$ bands and their bonding combination corresponds to the Zhang-Rice doublet. A consistent theoretical description requires a treatment of the correlated $2p$ and $3d$ bands on equal footing as has been demonstrated by Kune\v{s} \emph{et al.}~\cite{kunes_local_2007} using the local density approximation (LDA) plus dynamical mean field theory (DMFT) scheme.

\paragraph{Time-resolved two-photon photoemission}

The excitation across the optical band gap as studied in the present work is sketched in Fig.~\ref{2PPE}(a,b).
The photoexcitation transfers an electron into the UHB, resulting in a hole in the ligand $p$ orbital (a holon state) and a triply occupied site in the Ni $e_g$ manifold (a triplon state).
We explore these states and their ultrafast dynamics by the subsequent UV probe pulse, which photoemits the excited electron and projects the whole system to a final state with a ligand hole.
Pump-probe 2PPE spectra are shown in Fig.~\ref{2PPE}(c) and (d) for a NiO(001)-(1$\times$1) ultrathin film grown on Ag(001) at zero delay between the UV pulses.
The 2PPE spectra reveal one occupied state at $-1.3$\,eV with respect to the Fermi level $E_F$, marked as V in Fig.~\ref{2PPE}(b,d), and two unoccupied states at 2.0 and 2.5\,eV above $E_F$ marked as S and C, respectively, in Fig.~\ref{2PPE}(b,c).
The state V is at the top of the oxygen 2$p$ band and is assigned to the Zhang-Rice doublet ($3d^{8}Z^{-1}$) similarly to the one-hole final state in conventional photoemission experiments~\cite{bala_zhang-rice_1994,kunes_local_2007,taguchi_revisiting_2008}.
The states C and S mark the lower edge of the UHB and the well-known NiO(001) 3$d_{z^2}$ surface state \cite{kodderitzsch_exchange_2002,schron_influence_2013}, respectively.
Their energies depend only weakly on the NiO film thickness as summarized in Fig.~\ref{2PPE}(b) for epitaxial films with thicknesses between 4 and 20\,ML.
The size of the NiO charge-transfer gap can be estimated from the energy difference between the states V and C as $\Delta=(3.8\pm0.2)$\,eV in agreement with literature values~\cite{hufner_electronic_1994}.

\begin{figure}[t]
\includegraphics[width=0.9\linewidth]{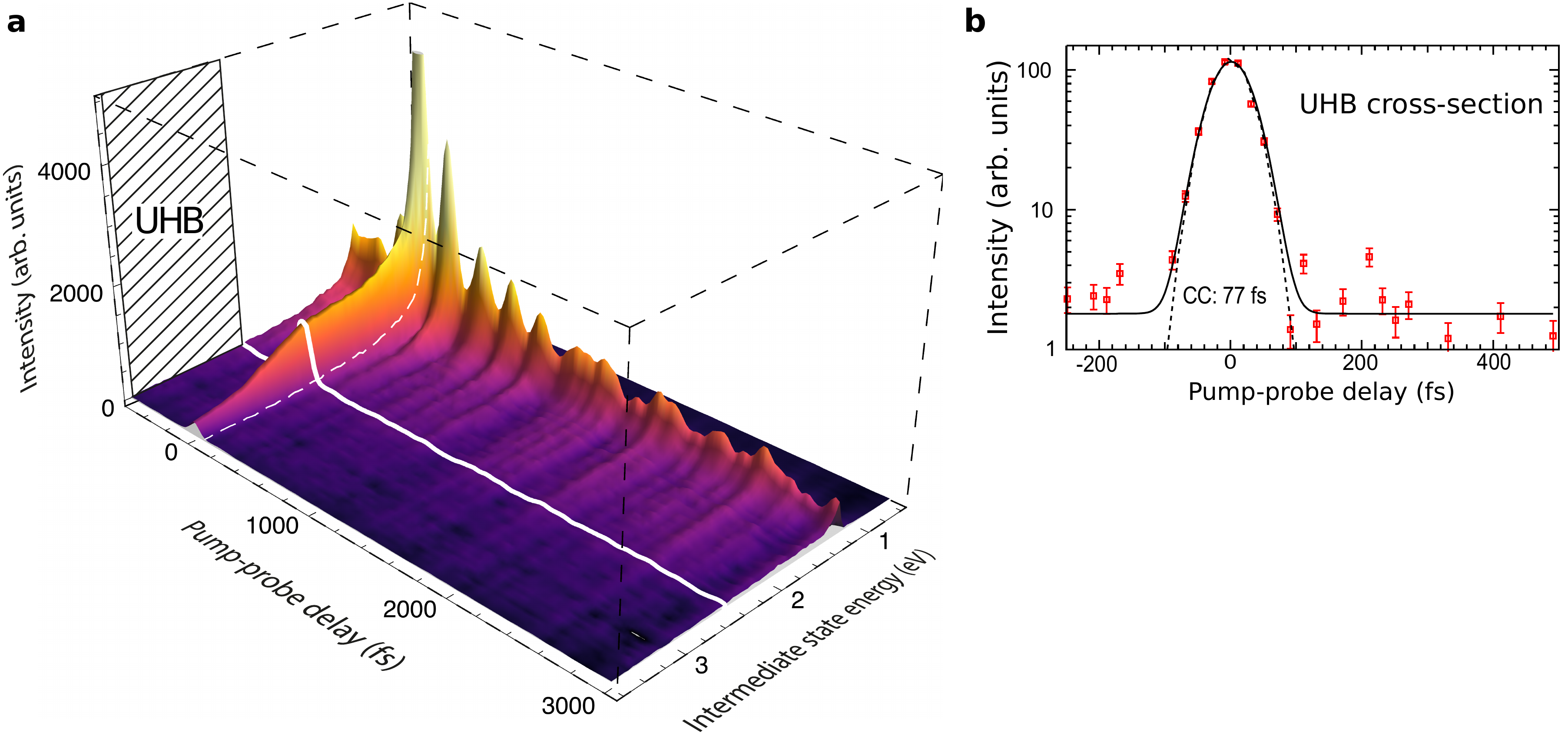}
\caption{Time-resolved pump-probe 2PPE data. \textbf{a}, Experimental data
  ($h\nu_\text{pump}=4.2$\,eV, $h\nu_\text{probe}=3.4$\,eV, T=150\,K) for a 9\,ML NiO(001) thin film with
  background subtracted and smoothing using a low-pass filter (without filtering in Fig.~\ref{temp}).
  The thick white line indicates the position of the lower edge of
  the UHB at 2.5\,eV. \textbf{b}, 2PPE intensity at the UHB bottom versus pump-probe
  delay. The cross-correlation (CC) of the pump and probe pulses (dashed line) has been determined
  from the 2PPE signal at higher energies.}
\label{yaroslav}
\end{figure}

The time-resolved 2PPE spectra are depicted in Fig.~\ref{yaroslav}(a) for pump-probe delays up to 3\,ps. The pump photon $h\nu_\text{pump}$\,=\,4.2\,eV promotes the electron into the UHB with a maximum excess energy of about 0.4~eV.
The relaxation within the UHB and the population of the in-gap state takes place on ultrafast time scales well below the temporal width of our pump-probe cross correlation, which has been measured to be about 80\,fs, see Fig.~\ref{yaroslav}(b).
A detailed analysis of the time traces around zero delay results in an upper limit of 10\,fs for the electron lifetime in the UHB (see Supplementary Information).
We emphasize that the observed electron lifetime at the bottom of the (undoped) UHB is three orders-of-magnitude shorter than the value expected at the conduction band minimum of conventional semiconductors with a similar band gap~\cite{tisdale_electron_2008,yukawa_electron-hole_2014,mickevicius_time-resolved_2005}.

\begin{figure}
  \includegraphics{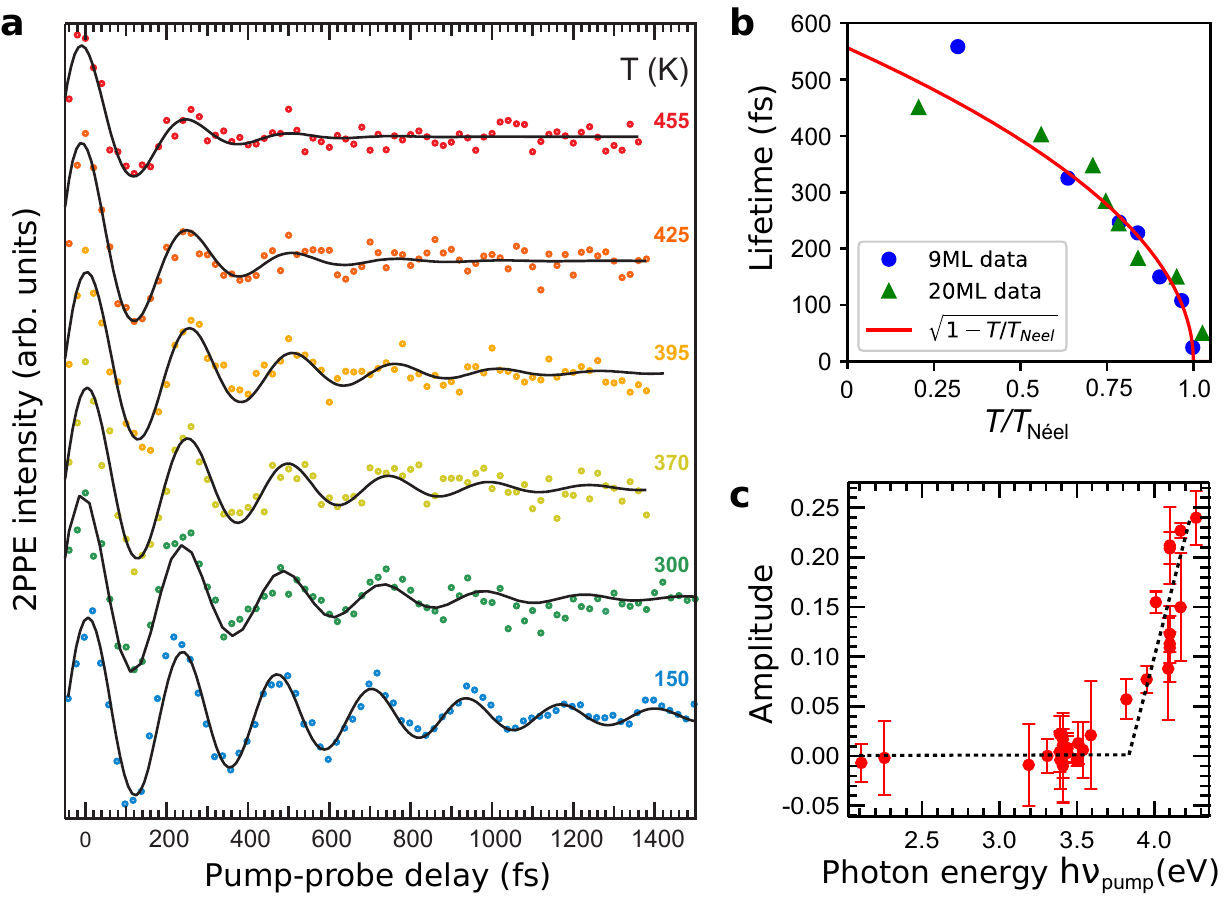}
 \caption{Oscillatory contribution of the 2PPE signal for the in-gap state:
   \textbf{a}, Time-dependent 2PPE signal of a 9\,ML
   NiO(001) thin film  for temperatures between 150 and 455\,K (1.2\,eV intermediate state energy,
   $h\nu_\text{pump}=4.2$\,eV and $h\nu_\text{probe}=3.4$ eV, slowly decaying
   background subtracted). The black solid lines describe exponentially damped
   harmonic oscillations. \textbf{b}, Oscillation lifetime $\tau_d$ extracted
   from the time-dependent 2PPE signal for NiO(001) thin films of 9 and 20\,ML.
   Note that $T_{N}$ of the 20\,ML film is reduced by about 60\,K for films of
   reduced thickness (9\,ML). \textbf{c}, Pump-photon energy dependence of the
   oscillation amplitude at $t=0$ extracted from the time-dependent 2PPE signal
   for the NiO(001) thin film of 9\,ML at 300\,K.
   }
\label{temp}
\end{figure}

In order to understand the efficient energy dissipation of the UHB electrons, we
take a closer look at the 2PPE spectra in the low-energy region. In
Fig.~\ref{yaroslav}, we observe a long-lived oscillatory contribution at an
intermediate state energy of 1.2\,eV, slightly above the low energy
photoemission cut-off. An intensity profile across this energy region reveals an
oscillatory component on top of a slowly decaying background (see
Fig.~\ref{yaroslav} as well as Fig.~\ref{temp}(a)). It is well described by a
damped oscillation $I_\text{osc}(t)\,=\,A\cdot e^{-t/\tau_d}\cdot
\cos(2\pi\nu_b\cdot t\,+\,\phi)$ with a frequency of
$\nu_b\,=\,(4.20\pm0.26$)\,THz and a dephasing time of $\tau_d=(558\pm35)$\,fs.
The oscillation frequency corresponds to $h \nu_b$\,=17\,meV.

Figure \ref{temp}(a) shows the oscillatory signal for the 9\,ML ultrathin film in a temperature range from 150 to 455\,K.
Whereas the oscillation amplitude at $t=0$ and the oscillation frequency do not vary significantly, the damping rate increases strongly at higher temperatures.
This temperature dependence is summarized in Fig.~\ref{temp}(b) together with the data for a thicker film of 20\,ML.
In both cases, we find long damping times at low temperatures that can be extrapolated to about 550\,fs at 0\,K.
The temperature dependence of the damping time follows the functional form $\sqrt{1-T/T_{N}}$ up to the \Neel temperature ($T_{N}$), where it approaches zero.
This dependence points to a dominant role of the coupling to the antiferromagnetic spin system and constitutes our main finding. Note that this dependence is valid for the 20\,ML as well as for the 9\,ML film, despite a 60\,K difference in their $T_{N}$.
For both thin films, $T_{N}$ has been determined experimentally by the disappearance of the antiferromagnetic ($2\times1$) superstructure seen in low-energy electron diffraction (LEED, see Supplementary Information).
As compared to NiO bulk, $T_{N}$ is reduced to 480\,K and 420\,K for 20 and 9\,ML thin films, respectively, in agreement with Ref.~\onlinecite{Alders98}.

Whereas the observed amplitude at $t=0$ is only weakly temperature dependent, it shows a strong threshold behavior with respect to the pump photon energy ($h\nu_\text{pump}$).
The $h\nu_\text{pump}$-dependence is exemplified in Fig.~\ref{temp}(c) for the 9\,ML film, with a clear onset at $h\nu_\text{pump}=3.8$ eV that agrees with the magnitude of the charge transfer energy $\Delta$.
This indicates that the THz oscillations are triggered by the excitation of an O 2$p$ electron across the charge transfer gap into the UHB. 
Therefore we will focus on the electron dynamics activated by the photoexcitation into the UHB in the following, while the lower $t_{2g}$ states will not be considered.
On the intermediate state energy scale, the THz oscillations have a significant amplitude from the photoemission onset up to 1.5\,eV above $E_F$ (see Supplementary Information), which implies an in-gap state located at around 1\,eV below the UHB as marked in Fig.~\ref{2PPE}(b) in light red.

\paragraph{Two-band t-J model}

To describe the observed electron dynamics in NiO, we developed a two-band extension of the $t$-$J$ model in the AFM phase. This model neglects excitonic effects and focuses on the triplons only, {\it i.e.}, we have projected out the local many-body states which are either unoccupied, singly or fully occupied.
The total Hamiltonian is composed of the local, kinetic and the superexchange parts, $H=H_\text{loc}+H_\text{kin}+H_\text{ex}.$
The local part captures the Hubbard and the Hund physics, which is described in Methods in detail.
The usual second-order perturbation theory leads to the AFM exchange Hamiltonian 
\begin{equation}
H_\text{ex} =J_\text{ex} \sum_{\alpha\<ij\>}\vS_{i\alpha}\cdot\vS_{j\alpha},
  \label{exchange}
\end{equation} 
where $\vS$ is a spin-$\frac12$ operator, $\langle i,j\rangle$ denotes nearest-neighbor pairs, and the subscript $\alpha$ labels 
different $e_g$ orbitals. The formation of AFM spin-order competes with the delocalization of triplon states, with the latter described by
\begin{equation}
 H_\text{kin} = 
 -t_0 \sum_{\langle i, j \rangle} \sum_{\alpha \sigma}(\cc_{i \alpha \sigma} \ca_{j \alpha \sigma} + \cc_{j \alpha \sigma} \ca_{i \alpha \sigma}).
  \label{kin}
\end{equation}
Here $t_0$ denotes the nearest-neighbor hopping and $\tilde c^\dagger$ the (projected) electron creation operator.
This competition is responsible for string-like states at the lower edge of the UHB~\cite{trugman_interaction_1988,dagotto_correlated_1994,obermeier_properties_2010,golez_mechanism_2014}.

In equilibrium, the ground state is dominated by the local high-spin states $S=1$ (S1 in Fig.~\ref{Denis2}) forming an AFM spin configuration.
The remaining local many-body states are illustrated in Fig.~\ref{Denis2} and include the triply occupied state (triplon, T), the minority high-spin states (S1$^{\flat}$) and two types of the low-spin $S=0$ Hund excitations.
In the first type, the two electrons occupy two different orbitals with an energy cost of $J_H$ (S0) relative to the ground state, whereas in the second type the same orbital is occupied by two electrons with an energy cost of $3J_H$ (S0$^{\flat}$).
These localized excitations have been the subject of intense experimental~\cite{fromme_spin-flip_1994,fiebig_second_2001,muller_angle-resolved_2008} and theoretical~\cite{mackrodt_bulk_2000,satitkovitchai_ab_2005,domingo_ab_2012} investigations.
In addition, a spin-flip excitation S1$^{\flat}$ at energy $2zJ_\text{ex}$ is taken into account, with $z$ denoting the coordination number of the neighboring TM ions coupled by the exchange interaction.

\begin{figure}
  \includegraphics[width=1.0\linewidth]{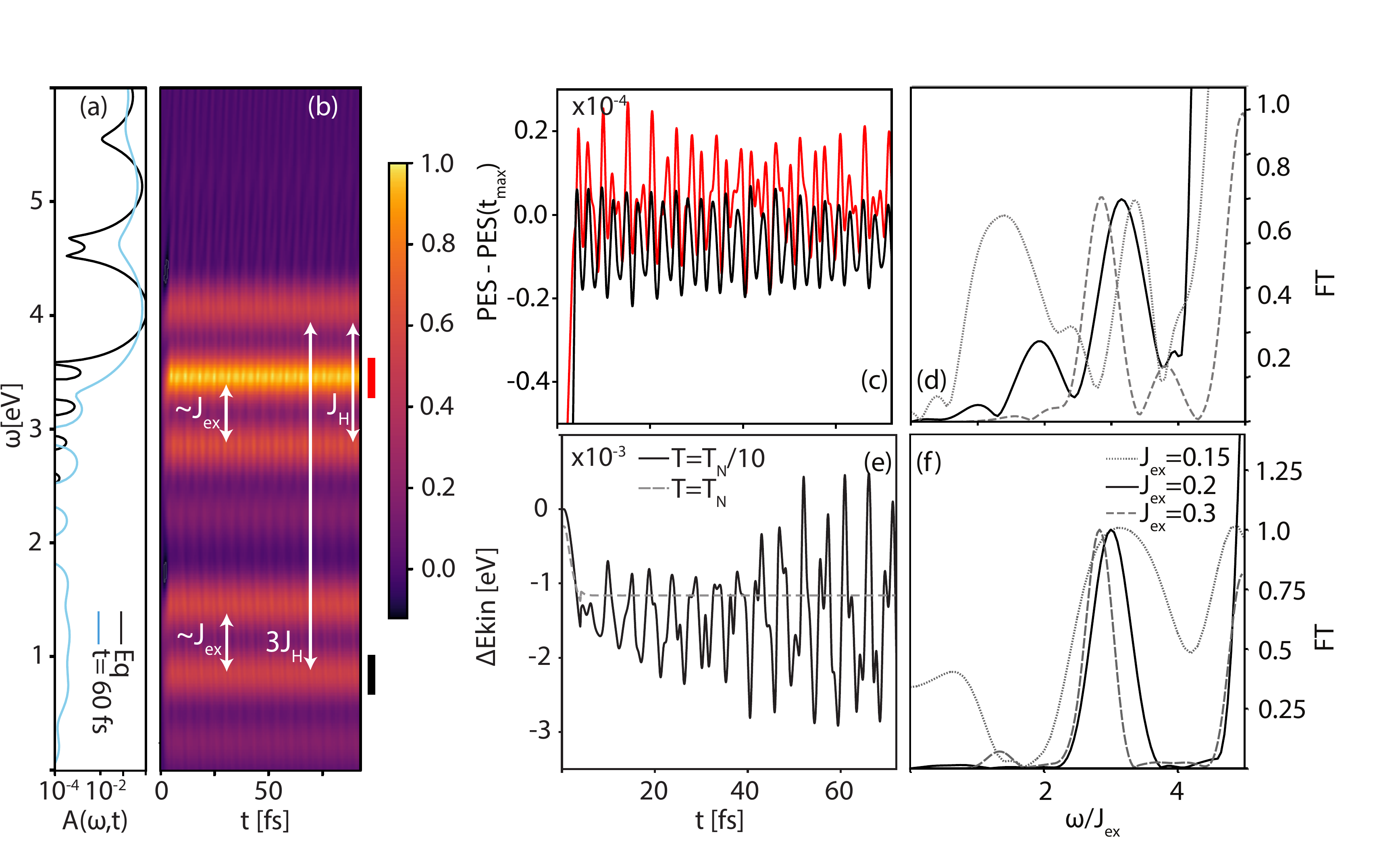}
  \caption{\textbf{a}, Spectral function $A(\omega,t)$ in equilibrium (black) and in the
    photoexcited state (light blue) for $J_\text{ex}=0.2$\,eV. \textbf{b}, Theoretical
    time-dependent photoemission spectrum (PES), see Eq.~\eqref{eq:PES}, after the
    photoexcitation, exhibiting coherent oscillations of the photo-induced in-gap
    signal. \textbf{c}, Time evolution of the energy-integrated PES in the window
    [3.4,3.8]\,eV (red) and [0.8,1.2]\,eV (black), see also the vertical color bars in
    panel \textbf{b}. \textbf{d} Normalized Fourier transform of the integrated PES signal
    for different values of the superexchange interaction $J_\text{ex}=0.15,0.2,0.3$
    eV. The background evolution has been subtracted using a low-order spline
    interpolation. \textbf{e}, Time evolution of the kinetic energy for a temperature
    below and equal to $T_{N}$. \textbf{f}, Fourier transform of the
    kinetic energy for different values of the superexchange interaction
    $J_\text{ex}=0.15,0.2,0.3$\,eV. The background has been subtracted using a low-order
    spline interpolation.}
\label{Denis1}
\end{figure}

In the calculated equilibrium spectral function (black line in Fig.~\ref{Denis1}(a)), sharp resonances at the lower edge of the UHB around 4 and 5\,eV can be identified.
These peaks are reminiscent of the string state features in the single band $t$-$J$ model \cite{trugman_interaction_1988,dagotto_correlated_1994,obermeier_properties_2010,golez_mechanism_2014}.
To model the pumping in experiments by the UV excitation across the charge-transfer gap $\Delta$, in the theoretical model the system is excited by photo-doping of electrons into the lower edge of the UHB.
During the photo-doping process, triplons $T$ are produced.
The photoexcited system is subsequently probed by photoemission in the experiments, which corresponds to the removal of an electron from the whole system.
Upon the removal of one electron from the triplon state, the system can end up in a low-spin S0 or S0$^\flat$ state, which produces the in-gap feature in the light blue spectrum in Fig.~\ref{Denis1}(a).
In addition, a substantial population of the photo-induced in-gap states is found in the calculated time-dependent photoemission spectra\cite{freericks_theoretical_2009} in Fig.~\ref{Denis1}(b) (see Eq.~\eqref{eq:PES} in Methods). 
There, the splitting of the Hund excitation S0~(S0$^{\flat}$) from the lower edge of the UHB is given by the Hund coupling $J_H$~($3J_H$).
This characteristic signature allows us to identify the dominant feature in the experiments in Fig.~\ref{yaroslav}(a) as the Hund excitation S0$^{\flat}$.
On top of the Hund excitations, we can identify clear sidebands in Fig.~\ref{Denis1}(b) whose splitting is proportional to the superexchange $J_\text{ex}$.
The mixture of the Hund and the AFM excitations as seen in Fig.~\ref{Denis1}(b) represents a non-equilibrium, non-thermal manifestation of the string states in multiband systems.
Moreover, all these in-gap features exhibit transient oscillations in agreement with the experimental findings, which will be discussed in the following.

The time-resolved photoemission intensity associated with the Hund excitations shows fast oscillations in Fig.~\ref{Denis1}(c), which correspond to the energy-integrated intensity in the red and black energy windows in Fig.~\ref{Denis1}(b).
On top of these fast oscillations, a slower oscillation at a characteristic frequency proportional to the superexchange $J_\text{ex}$ can be identified in the black curve in Fig.~\ref{Denis1}(c) as well as in the Fourier-transformed spectra in Fig.~\ref{Denis1}(d). 
Since the slow oscillation scales clearly with $J_\text{ex}$ as shown in Fig.~\ref{Denis1}(d), we attribute its origin to the AFM state of NiO.
To further support this interpretation, we compare the time evolution of the kinetic energy below and at $T_{N}$ in Fig.~\ref{Denis1}(e).
Indeed, the long-time coherent oscillation only appears at temperatures below $T_{N}$.
Note that an artificially large value of the exchange interaction $J_\text{ex}$=200\,meV has been used in order to observe the slow magnetic dynamics on the time scales accessible in the simulations.
The realistic value according to Dietz \emph{et al.}~\cite{dietz_infrared_1971} and Hutchings and Samuelsen~\cite{hutchings_measurement_1972} is $J_\text{ex}\approx 19$\,meV for the next-nearest neighbor coupling $J_2$.
The nearest neighbor ferromagnetic exchange coupling $J_1=-1.37$\,meV is one order-of-magnitude smaller and is neglected in our modelling.
The frequency of the coherent AFM oscillations in Fig.~\ref{Denis1}(d) and (f) scales approximately linearly with the superexchange $J_\text{ex}$.
Typically, the frequency of these oscillations is $\omega/J_\text{ex}\approx3$ roughly matching the distance between the equilibrium string
states, see Fig.~\ref{Denis1}(a).
We anticipate that for realistic superexchange interactions the string sidebands associated with a given Hund excitation in Fig.~\ref{Denis1}(b) merge within the experimental resolution.

The initial fast relaxation of the photo-excited triplons happens on a time scale of 10\,fs and results from the scattering of  triplons with the high-spin states, producing Hund excitations $S=0$.
Since the Hund excitations are mainly determined by the local $J_{H}$, they are little influenced by the AFM background.
As a result, the initial relaxation process in Fig.~\ref{Denis1}(e) is similar below and above $T_{N}$. 
The initial relaxation is strongly enhanced for high-frequency excitations, where photoexcited electrons are injected up to the upper edge of the UHB (see Supplementary Information).
In these processes, the local spin configurations can absorb energy quanta of $J_H$ or $3J_H$ on the time scale of the inverse hopping amplitude $1/t_0$, which is a very efficient dissipation mechanism.
This relaxation channel closes once the excess kinetic energy drops below $J_H$, since this is the minimum energy that can be dissipated by local spin excitations.
This is the Hund's equivalent of the phonon-window effect~\cite{sentef_examining_2013,rameau_energy_2016,golez_relaxation_2012}.

\begin{figure}[h!]
\includegraphics[width=1.0\linewidth]{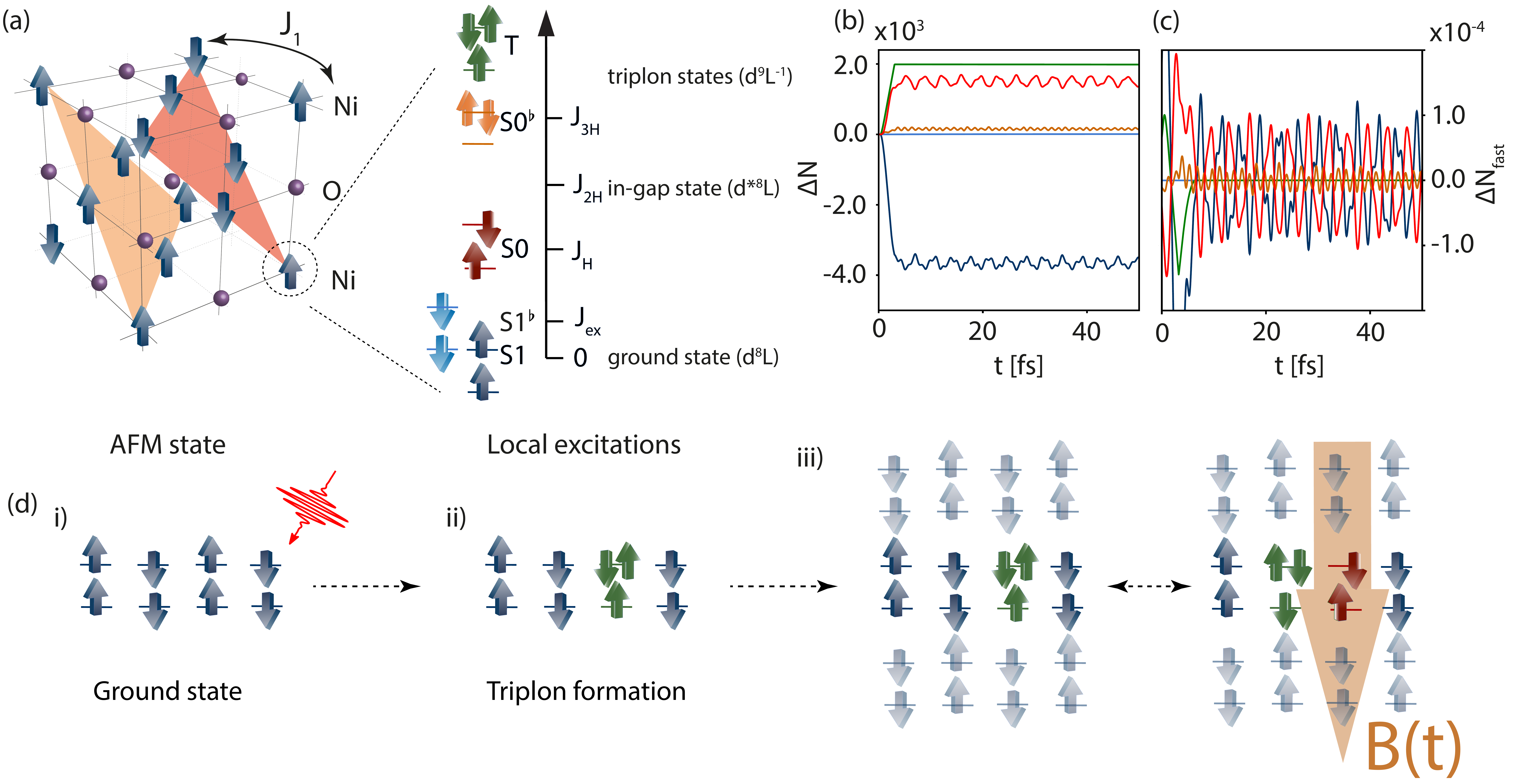}
\caption{\textbf{a}, Schematic view of the NiO lattice structure and relevant local many
  body states. The latter are the triplon~(T), majority high-spin doublon~(S1), minority
  high-spin doublon (S1$^{\flat}$), low-spin Hund excitation with electrons in different
  orbitals at an energy cost $J_H$ (S0), and low-spin Hund excitation with electrons in
  the same orbital at an energy cost $3J_H$ (S0$^{\flat}$). Spin-flip excitations such as
  S1$^{\flat}$ are described by the exchange Hamiltonian~\eqref{exchange}. \textbf{b},
  Time evolution of the occupation for the most relevant local many-body states on the A
  sub-lattice, and \textbf{c}, the same data with subtracted background dynamics. The
  colors of the curves match the graphical representation of the local many-body states.
   \textbf{d}, Schematic view of the photo-induced dynamics in NiO: (i) AFM ground
  state with high-spin~(S1) states, (ii) photo-induced state with mobile triplons, (iii)
  coherent dynamics between two many-body states, where the magnetic background $B(t)$ is
  responsible for the characteristic oscillations whose frequency scales with the strength of the superexchange
  interaction $J_\text{ex}$.
  }
\label{Denis2}
\end{figure}

\paragraph{Discussion}
Coherent electron dynamics has been observed in isolated quantum systems with few relevant energy levels~\cite{kinoshita_quantum_2006} or for image potential states~\cite{hofer_time-resolved_1997,marks_quantum-beat_2011}.
While these phenomena are related to well isolated quasiparticle physics~\cite{schuler_time-dependent_2016}, coherent many-body phenomena are more exotic. Recently, such an observation has been reported in a large array of trapped cold atoms~\cite{bernien_probing_2017}.
However, our work provides the first observation of coherent long-lived many-body dynamics due to coupling to the AFM background in a solid.
Similar phenomena have been detected in cuprates exhibiting oscillations with a very short lifetime~\cite{miyamoto_probing_2018}, but their origin remains controversial~\cite{bittner_coupled_2018,shinjo2018}.

The experimental observations raise the central question how a strongly correlated solid can support coherent oscillations on such a long time scale (up to 2\,ps).
In order to avoid dephasing due to the coupling with a continuum of states, there must exist a small subset of many-body states that are well isolated.
The theoretical analysis in Fig.~\ref{Denis2}(b) allows us to follow the occupation of the local many-body states in the time evolution.
After the photoexcitation, the majority high-spin states (dark blue) are depopulated at the expense of low-spin Hund excitations.
The subsequent dynamics is governed by coherent oscillations between the high- and the low-spin states, which can be more clearly seen in Fig.~\ref{Denis2}(c) where the slow background has been subtracted. Since the kinetic energy of the triplons is too small to produce Hund excitations~($E_{\text{kin}}<J_H$), the triplons are trapped between nearest-neighbor sites as exemplarily shown in Fig.~\ref{Denis2}(d)~(iii). This also explains why the in-gap states in the PES are isolated. An analogous process is possible for the high-energy Hund excitations S0$^{\flat}$ as well.
Furthermore, each creation of a low-spin excitation is accompanied by a ferromagnetic (FM) disturbance in the spin background, which produces a feedback on the spin system in the form of an effective magnetic field $B(t)$ and induces the coherent oscillations at a frequency controlled by the superexchange interaction $J_\text{ex}$.

The lifetime of the coherent oscillations is determined by their decay channels.
A higher-order process leads to the formation of a FM domain wall and results in a steady increase in the population of the minority high-spin configuration S1$^{\flat}$.
The latter process can be further enhanced by high-frequency excitations up to the upper edge of the UHB (see Supplementary Information). 
This formation of FM strings is one possible relaxation process.
However, being a higher-order process with rather low probability, it cannot be responsible alone for the decay of the THz oscillations with a time constant of about $\tau_d=500$\,fs as observed in the experiments.
A spreading of localized FM excitations via magnons can provide an alternative decay channel which is not captured by the approximate theoretical description considered here and would be an interesting subject for future investigations. 
The binding of singlons and triplons and its impact on the recombination rate is not properly captured in DMFT, but should not play an important role in the dilute photoexcitation limit explored in the experiments.

In conclusion, we have combined the two-photon photoemission experiments and a two-band $t$-$J$ model to investigate the evolution of ultrafast photo-induced in-gap states in strongly correlated NiO.
Long-lived coherent oscillations at THz frequencies are observed in the experiments and identified in the theoretical modeling as a signature of the coupling between long-range antiferromagnetism and local Hund excitations. 
Contrary to common belief, this work shows that many-body coherent dynamics can be observed in solid-state systems despite the intrinsic coupling to numerous active degrees of freedom.
Our findings indicate an important interplay between local electronic excitations and the long-range ordering. The combined experimental and theoretical results pave the way to systematic explorations of the many-body physics in correlated oxides driven out of equilibrium by ultrafast above-band-gap excitations. 
An obvious extension of our work would be the investigation of similar multi-band effects in cold-atom experiments,
where two-dimensional antiferromagnetism has been realized recently~\cite{mazurenko2017}.

% #######################################################################################
\section{Theoretical Methods}
To describe the photo-doped state we project out the local many-body states which are either unoccupied, singly or fully occupied by introducing the projected operators $\tilde c_{i\alpha\sigma}=P^{\dagger} c_{i\alpha\sigma} P,$ where $c_{i\alpha\sigma}$ is the annihilation operator for site $i$, orbital $\alpha$ and spin $\sigma$ and $P$ is the local projector.
The total Hamiltonian is composed of the local, kinetic (Eq.~\eqref{kin}) and superexchange (Eq.~\eqref{exchange}) parts,  $H=H_\text{loc}+H_\text{kin}+H_\text{ex}.$ The local interaction $H_{\text{loc}}$ reads
\begin{multline}
 {H}_{\text{loc}} = 
 U \sum_{i,\alpha} \n_{i, \alpha \uparrow} \n_{i \alpha \downarrow} - \mu \sum_{i\alpha\sigma}\n_{i\alpha\sigma}
 + \sum_{i,\alpha < \beta} \sum_{\sigma, \sigma'}
   (U' - J_H\delta_{\sigma \sigma'}) \n_{i \alpha \sigma} \n_{i \beta \sigma'} \\
 + \gamma J_H \sum_{i,\alpha < \beta} \left(
 \cc_{i \alpha \uparrow} \cc_{i \alpha \downarrow} \ca_{i \beta \downarrow} \ca_{i \beta \uparrow}
 +
 \cc_{i \alpha \uparrow} \cc_{i \beta \downarrow} \ca_{i \alpha \downarrow} \ca_{i \beta \uparrow}
 \right)
 \, ,
 \label{eq:HlocHund}
 \end{multline}
where $U$, $U'$, and $J_H$ is the intra-orbital Coulomb, inter-orbital Coulomb, and the Hund's exchange interaction, respectively.
The interaction in Eq.~(\ref{eq:HlocHund}) is the Kanamori interaction, which becomes rotationally invariant when $U' = U - 2J_H$ and $\gamma = 1$. Here we set $U'= U-2J_H$ but employ the non-symmetric form (with $\gamma = 0$) for practical reasons.
We have assumed no crystal-field splitting between the two $e_g$ orbitals.
The half-filling condition is imposed by adjusting the chemical potential to $\mu=(3U-5J_H)/2$. The lattice of NiO is composed of a pair of interpenetrating magnetic sublattices, where one sublattice is coupled with a weak FM interaction and the second one by a much stronger AFM interaction~\cite{hutchings_measurement_1972}.
We neglect the presence of the weakly coupled FM sublattice and focus only on the AFM sublattice.
The AFM superexchange interaction is mediated via the $p$ orbital and modeled by the Heisenberg Hamiltonian~\eqref{exchange}.

To solve the electron dynamics, we use the non-equilibrium dynamical mean field theory~\cite{aoki_nonequilibrium_2014}, which maps a correlated lattice problem onto a self-consistently determined impurity problem.
While recent theoretical developments have enabled the description of photo-doped charge-transfer insulators, state-of-the-art simulations are still limited to the paramagnetic phase and short-time dynamics~\cite{golez2019multi,golevz2018B,tancogne2018}.
In order to simplify the description, we treat the impurity problem using the projection onto the subspace of relevant local many-body states. In addition, we use the lowest-order strong coupling expansion, the so-called non-crossing approximation~(NCA)~\cite{eckstein_nonequilibrium_2010,haule_anderson_2001,bittner_coupled_2018}.
In order to get access to the long-time behavior we solve the problem on a Bethe lattice, which does not compromise the analysis of the proposed mechanism as long as the realistic system is not one-dimensional, i.e. as long as spin-charge separation does not play a role.
We explicitly break the AFM symmetry and treat the feedback of the AFM order on the mean-field level \cite{obermeier_properties_2010} as $J_\text{ex}\sum_{ij} \vS_{i\alpha} \vS_{j\alpha}\rightarrow 3 J_\text{ex} \sum_{ij} \vec S_{i\alpha}^z \langle \vec S_{j\alpha}^z\rangle$ for both $e_g$ orbitals 
assuming the spin polarization in the $z$ direction.
This decoupling becomes exact in the limit of large connectivity.
The model parameters other than $J_\text{ex}$ are taken from equilibrium LDA+U\cite{anisimov_band_1991} and LDA+DMFT\cite{kunes_local_2007,zhang_dft+dmft_2017B} studies, which used $U\approx 8$\,eV, $J_H \approx 1.0$\,eV.
We infer the effective hopping $t_0\approx 1$\,eV from the bandwidth of the UHB in the calculation of Kune\v{s} \emph{et al.}~\cite{kunes_local_2007} and from the bandwidth of the $d^9$ state in the Bremsstrahlung isochromate spectroscopy~\cite{hufner_electronic_1994}.
The hopping $t_0$ is of indirect nature and is mediated by the hopping through the oxygen $2p$ states as explained by Zhang and Rice in the case of cuprates~\cite{zhang_effective_1988}.

In order to model the excitation from a deep fully occupied $p$-band to the UHB we approximate the excitation by a sudden electron doping in a certain energy range of the UHB.
This is justified, since the time scale for the UV excitation at $h\nu$ = 4.1\,eV corresponds to 0.1\,fs according to  Heisenberg's uncertainty principle.
In addition, the fully occupied $p$-band is weakly interacting and can be integrated out.
The UV pump pulse may thus be modeled within the rotating wave approximation.
In all simulations we have attached the full bath at the lower edge of the UHB in the energy range $\omega=[3,4]$\,eV for the time interval [0, 2] \,fs.
The time-dependent spectral function $A(\omega,t)$ is obtained from the retarded component of the local Green's function $G^R$ as
\bsplit{
  A(\omega, t)=-\frac{1}{\pi}\mathrm{Im}\int_t^{t+t_{\text{cut}}} d{t'} e^{\I\omega(t'-t)} G^R(t', t).
}
and the photo-excitation spectrum is obtained from the lesser component $G^<$ as 
\bsplit{
\label{eq:PES}
	&\text{PES}(\omega, t)=\frac{1}{ \pi } \mathrm{Im}\int_t^{t+t_{\text{cut}}}  d{t'} e^{\I\omega(t'-t)} G^< (t', t).
}
Due to the limited propagation time, we have employed a forward integration for the Fourier transform in contrast to Ref.~\onlinecite{freericks_theoretical_2009}, where the Fourier integral has been done in relative time. 

\section{Experimental Methods}

The experiments have been performed in an ultrahigh vacuum chamber equipped with a 150\,mm hemispherical electron analyzer (Phoibos 150, SPECS, Berlin) with a 2D CCD detector and low-energy electron diffraction optics as described in detail elsewhere \cite{duncker_momentum-resolved_2012,gillmeister_image_2018}.
A broadly tunable femtosecond laser system operated at a repetition rate of 1.4\,MHz with two noncollinear optical parametric amplifiers (NOPA), pumped by a 20\,W fiber laser (IMPULSE, Clark-MXR, Dexter) is used for the 2PPE \cite{hofer_laser-excited_2011,duncker_momentum-resolved_2012}.
The two frequency-doubled beams of the NOPAs serve as pump-probe excitation source with typical pump-probe cross correlation widths between 70 and 100\,fs (full-width-at-half-maximum).

NiO thin films have been prepared by reactive Ni deposition at room temperature on a Ag(001) single crystal in 10$^{-6}$\,mbar O$_2$. Subsequent annealing is applied to achieve long-range order as monitored by LEED and high-resolution electron energy loss spectroscopy \cite{kostov_surface-phonon_2013,Kostov16b}. 

\section{Data availability}
The data that support the plots and other findings presented in this paper are available from the corresponding author upon reasonable request.

\section{Competing interests}
The authors declare no competing interests.

\section{Author contributions}
K.G. and W.W. designed the experiments. K.G. prepared the samples and measured
the 2PPE data. The data analysis and evaluation was done by K.G., C.-T.C., Y.P., and
W.W.. D.G., N.B., Y.P and P.W. constructed the model description and D.G. carried out
the DMFT calculations. All authors discussed the results and co-wrote the paper.

\begin{acknowledgments}
 The calculations have been performed on the Beo04 cluster at the University of
 Fribourg and on the Rusty cluster at the Flatiron Institute using a software
 library developed by M. Eckstein and H. U. R. Strand. The Flatiron Institute is a division
 of the Simons Foundation. K. G., C.-T. C., and W. W. acknowledge financial
 support from the German Research Foundation (Deutsche Forschungsgemeinschaft,
 DFG) through SFB 762 (A3, B8) and SFB/TRR 227 (A06). D. G., N. B. and P. W.
 were supported by Swiss National Science Foundation grant 200021-165539 and the
 European Research Council through ERC consolidator grant No. 724103. Y. P.
 acknowledges support of the DFG via SFB/TRR 173. D. G. would like to acknowledge 
 R. Sesny for help with graphical representation.
\end{acknowledgments}


% \bibliography{MyLibrary}

\newpage
\textbf{\large{Supplementary Information: Ultrafast coupled charge and spin dynamics in strongly correlated NiO}}

\maketitle
\paragraph{Magnetic characterization}

\begin{figure}[h!]
  \includegraphics[width=0.85\linewidth]{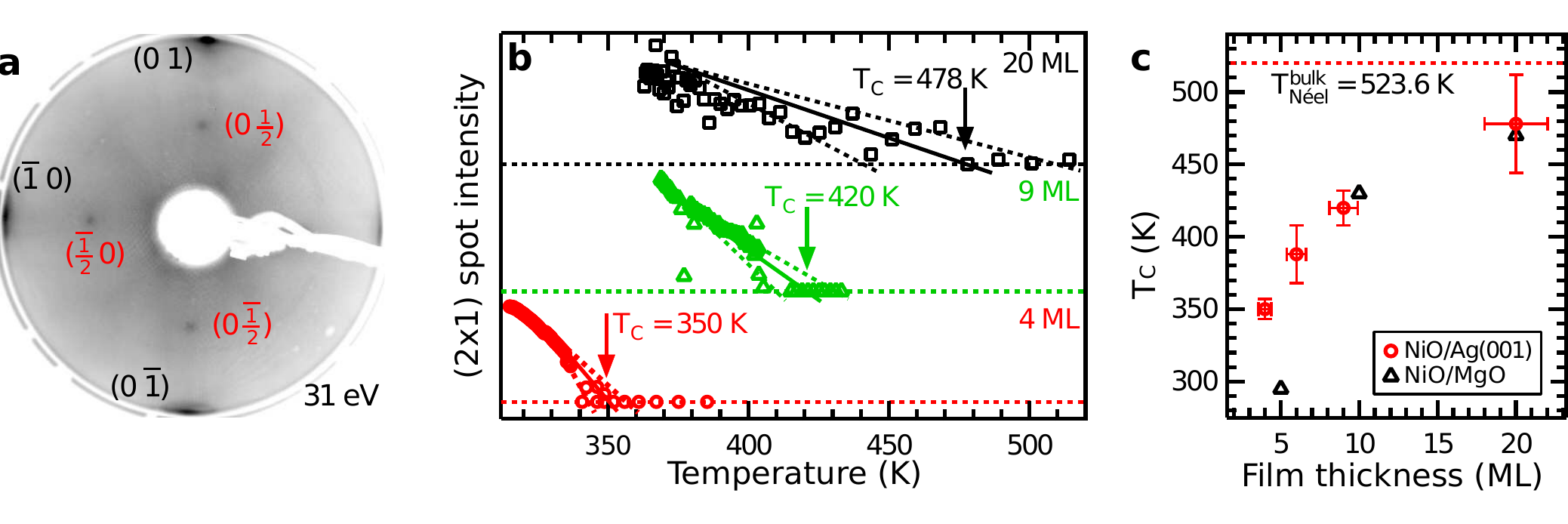}
 \caption{\textbf{a}, LEED pattern of a 4\,ML NiO(001) thin film on Ag(001) for
 an electron energy of 31\,eV. At this energy, the half-order spots due to the
 magnetic (2$\times$1) superstructure (marked in red) are clearly visible. Note that the first
 order diffraction spots are located at the rim of the LEED screen. 
 \textbf{b}, Temperature-dependent (2$\times$1) intensities for NiO films with 4, 9 and 20\,ML thickness. 
 The \Neel temperature is determined by the vanishing magnetic signal as indicated. 
 \textbf{c}, Extracted \Neel temperatures versus NiO film thickness (red open circles). Literature 
 data from Ref.~\onlinecite{Alders98} for NiO films on MgO(001) are marked by black triangles.}
\label{FigLEED}
\end{figure}

The NiO(001) thin films have been prepared by reactive metal deposition in an
oxygen atmosphere. The films have been characterized by low-energy electron diffraction (LEED) for a sharp
NiO(001)-(1x1) pattern. To assure a long-range antiferromagnetic order, the
weaker (2x1) superstructure due to the doubled magnetic unit cell~\cite{Palmberg68} have been analyzed
as depicted for a 4\,ML film at room temperature in Fig.~\ref{FigLEED}(a). At
a kinetic energy of 31\,eV, the weak but sharp half-order spots are clearly
discernible. The integrated half-order spot intensity is strongly temperature
dependent and vanishes at the \Neel temperature. This is depicted in
Fig.~\ref{FigLEED}(b) for NiO(001) films with 4, 9, and 20\,ML thickness. The
\Neel temperature as determined from the vanishing magnetic (2x1) signal is
shown in Fig.~\ref{FigLEED}(c) as red circles for all investigated film
thicknesses. In addition, data for NiO films on MgO(001) from Ref.~\onlinecite{Alders98}
are marked by black triangles. In agreement with earlier
studies, we find that the \Neel temperature is reduced for ultrathin films as
compared to $T_c= 523.6$\,K for NiO bulk~\cite{Marynowski99,altieri_image_2009}.

\paragraph{Time-resolved two-photon photoemission (2PPE)}

\begin{figure}
\includegraphics[width=0.65\linewidth]{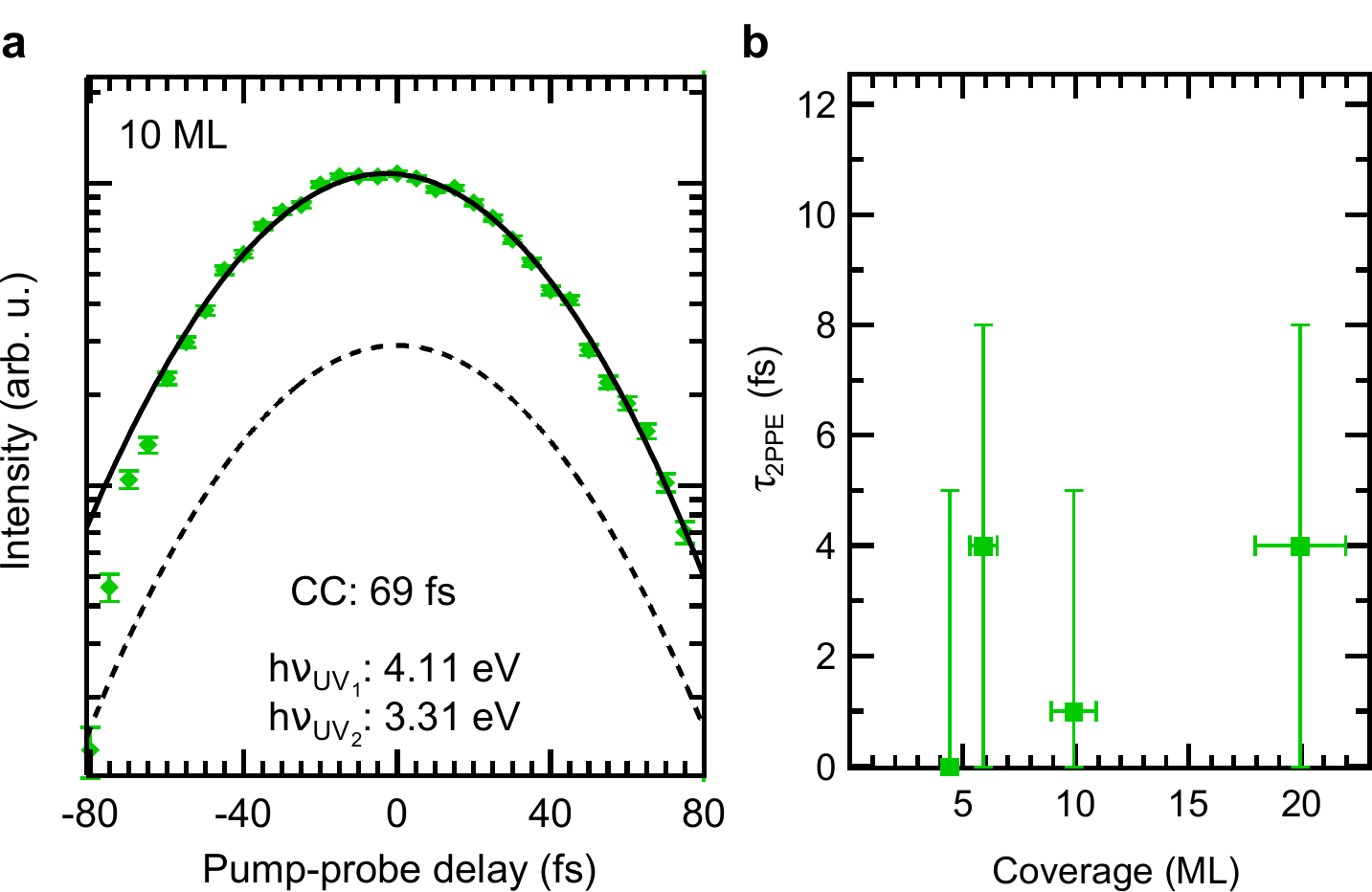}
\caption{\textbf{a}, Time-dependent 2PPE intensity 2.5\,eV above E$_F$ (bottom of the upper Hubbard band) as a
function of the pump-probe delay for a 10\,ML thick NiO film on Ag(001). 
The cross correlation of the pump and probe pulses (dashed line) has been determined from the 
2PPE signals at higher energies.
\textbf{b}, Extracted lifetime of the 2PPE signal at the bottom of the UHB for 
NiO thin films between 4 and 20\,ML thickness. In all cases, the lifetime is below 10\,fs.}
\label{FigLifetime}
\end{figure}

The time-dependent 2PPE signal at the bottom of the UHB (2.5\,eV above E$_F$) 
is depicted in Fig.~\ref{FigLifetime}(a) for a 9\,ML thick NiO(001) film on a
logarithmic intensity scale. Here pump and probe photon energies of 4.11 and 3.31\,eV, respectively, have
been used. The time-dependent signal follows directly the experimental cross
correlation trace, which is shown as dashed curve and has a FWHM of 69\,fs. We
estimate from this observation that the signal decays within a time shorter than
10\,fs. Similar fast decays have been found for all investigated film thicknesses as
summarized in Fig.~\ref{FigLifetime}(b).

\begin{figure}
  \includegraphics[width=0.45\linewidth]{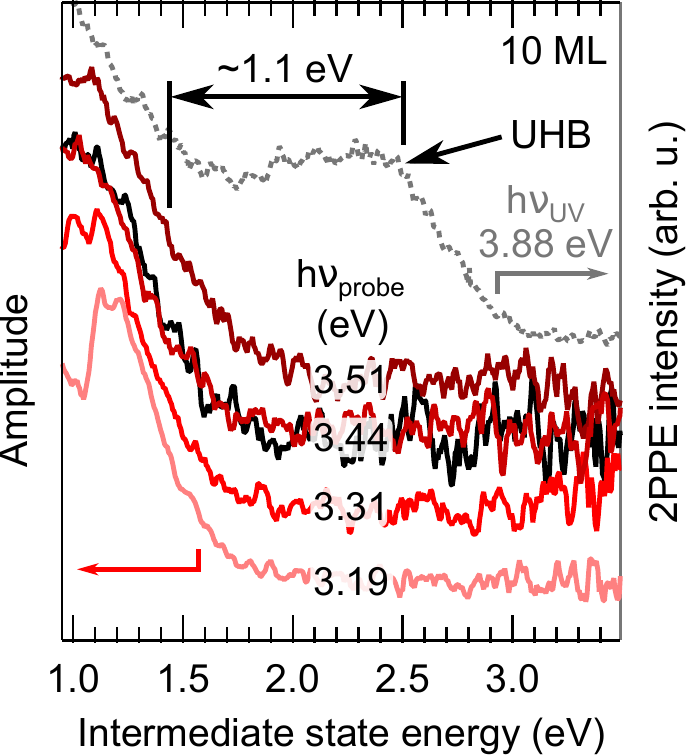}
  \caption{Amplitude of the 2PPE oscillations as a function of the intermediate state energy 
  after pumping with
  $h\nu_\text{pump}=4.11$\,eV (light red curves) and $h\nu_\text{pump}=3.95$\,eV (black
  curve). The probe photon energy $h\nu_\text{probe}$ is varied between 3.19 and 3.51\,eV
  as indicated in the graph. The grey photoemission spectrum is obtained by
  monochromatic 2PPE ($h\nu_\text{pump}=h\nu_\text{probe}=3.88$\,eV) and is provided to
  indicate the position of the UHB. Note that the THz oscillations start about 1\,eV
  below the bottom of the UHB. The spectra are shifted vertically for clarity.}
\label{FigGapState}
\end{figure}

\begin{figure}
  \includegraphics[width=0.5\linewidth]{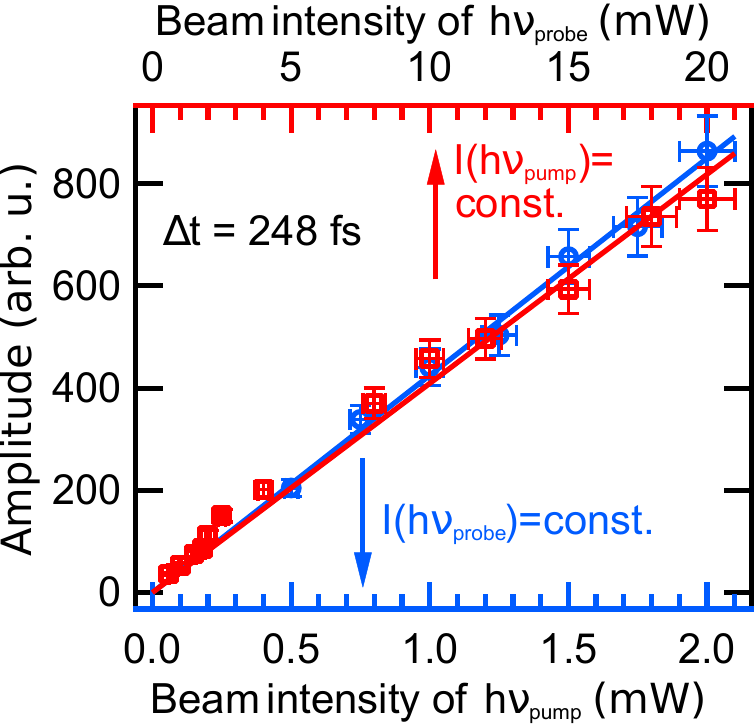}
 \caption{Amplitude of 2PPE oscillations as a function of the laser pump ($h\nu_\text{pump}=4.17$\,eV) and probe 
 ($h\nu_\text{probe}=3.39$\,eV) intensities for a 9\,ML thick NiO film on Ag(001). The time delay between the pump 
 and probe beams is fixed to 248\,fs. The oscillation amplitude scales linearly with both intensities. }
\label{FigIntensities}
\end{figure}

The 2PPE spectrum for the 10\,ML NiO film is depicted in Fig.~\ref{FigGapState}
as a gray trace for a monochromatic excitation with $h\nu_\text{pump}$ = $h\nu_\text{probe}$ =
3.88\,eV. As discussed in the main manuscript, excitations across the
charge-transfer gap lead to long-lived in-gap states that are located about 1\,eV
below the UHB and exhibit characteristic oscillations in their photoemission
intensities. The amplitude of these oscillations is shown in Fig.~\ref{FigGapState} 
(red traces, left axis) as a function
of the intermediate state energy for four different photon energies. The 2PPE 
oscillation amplitude is determined around a pump-probe delay of $\Delta$t= 248\,fs, 
which corresponds to the first oscillation maximum after time zero by the relative 
2PPE intensity variation (I$_\text{max}$-I$_\text{min}$)/I$_\text{min}$.
Figure~\ref{FigGapState} shows that in all cases the oscillations are observable
in the intermediate state energy range between 1 and 1.5\,eV above E$_F$, independent of the
final state energy. The dependence of this oscillation amplitude on the laser
fluence is depicted in Fig.~\ref{FigIntensities} for a 9\,ML NiO(001) thin film.
For a fixed pump fluence the amplitude scales proportionally
with the probe fluence (red squares). For a fixed probe fluence, it scales
proportionally with the pump fluence as marked by the blue circles.

\paragraph{High-frequency excitation}

\begin{figure}
  \includegraphics[width=1.0\linewidth]{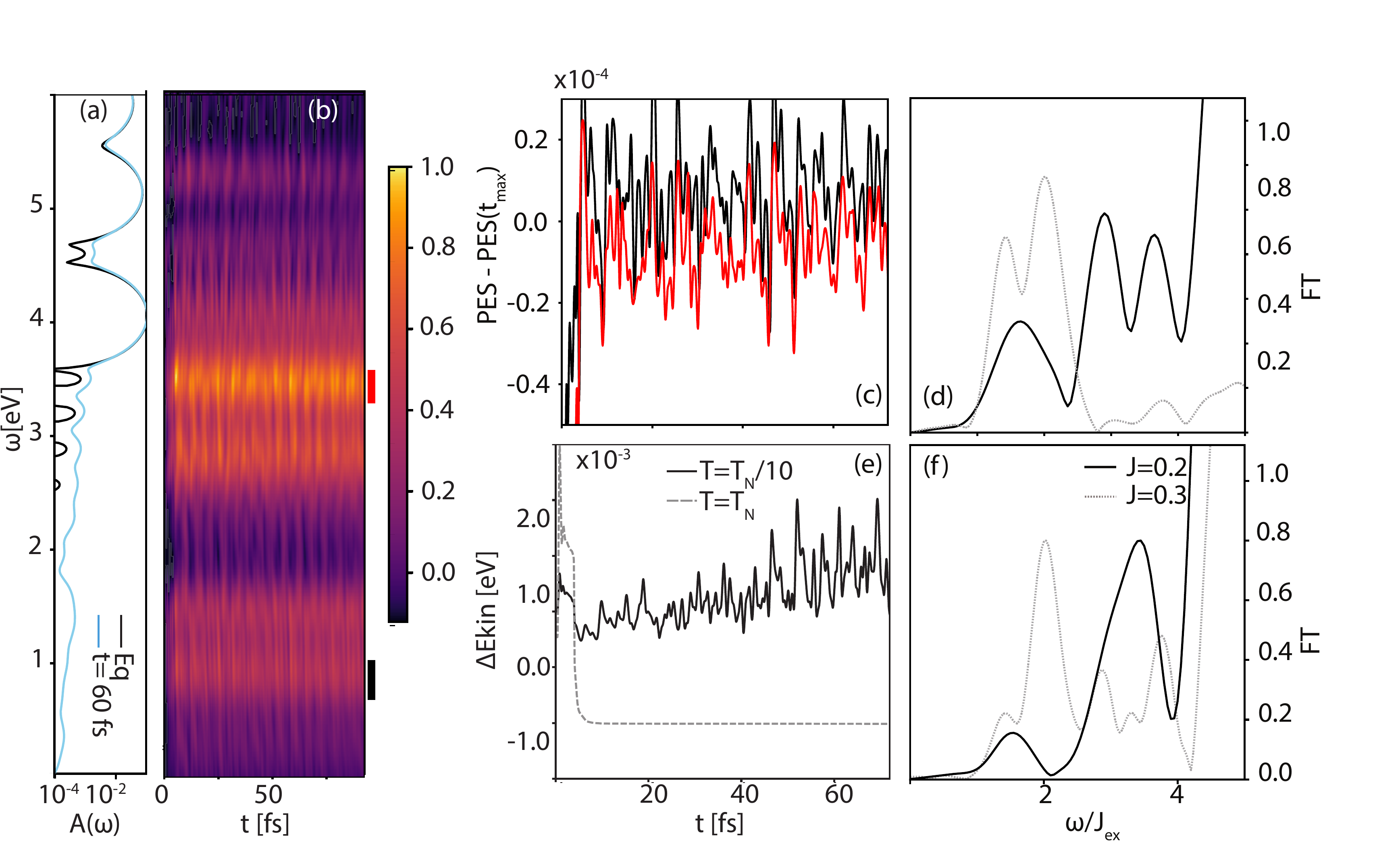}
  \caption{Photoexcitation with a high-frequency pulse, where electrons are injected in the energy window [8-10]\,eV: \textbf{a}, Spectral function in equilibrium (black) and in the photoexcited
    state  (light blue) for $J_\text{ex}=0.2$\,eV. \textbf{b}, Theoretical time-dependent photoemission spectrum (PES) after the photoexcitation
    showing the coherent oscillations of the photo-induced in-gap signal. \textbf{c}, Time evolution of the energy-integrated PES in the
    window [3.4,3.8]\,eV (red) and [0.8,1.2]\,eV (black), see also the vertical color bars in
    panel \textbf{b}. \textbf{d} Fourier transform of the integrated PES signal  for different values of the superexchange interaction $J_\text{ex}=0.15,0.2,0.3$\,eV, where the
    background evolution has been subtracted using a low-order spline
    interpolation. \textbf{e}, Time evolution of the kinetic energy for a temperature
    below and equal to the N\'{e}el temperature.  \textbf{f}, Fourier transform of the kinetic
    energy for different values of the superexchange interaction $J_\text{ex}=0.15,0.2,0.3$\,eV, where the background has been subtracted using a low-order spline
    interpolation.
    }
\label{SM:theory}
\end{figure}

Here, we simulate a high-frequency excitation injecting electrons at the top of the UHB in the energy window [8-10] eV using the same model parameters as in the main text. The electronic population created by the high-frequency pulse rapidly relaxes to the lower edge of the UHB, see Fig.~\ref{SM:theory}(b). The excess kinetic energy is absorbed by local Hund excitations leading to a finite population of the $S0$ states~\cite{strand_nonequilibrium_2017}. As the Hund excitations are gapped and the electrons can only absorb relatively large quanta of energy, the dynamics slows down after the excess kinetic energy becomes smaller than the Hund coupling, $\Delta E_\text{kin}<J_H,$ which is the Hund's physics equivalent to the phonon window effect~\cite{sentef_examining_2013,rameau_energy_2016,golez_relaxation_2012}. After the initial relaxation the occupation of the in-gap states and the kinetic energy show coherent oscillations, see Fig.~\ref{SM:theory}(c) and (e). Similar to the low-frequency excitation case presented in the main text, in Fig.~\ref{SM:theory}(e) these oscillations only exist below the \Neel temperature $T<T_N$. The Fourier analysis of these oscillations in Fig.~\ref{SM:theory}(d) and (f) shows that while slow oscillations do exist, they do not exhibit a clear scaling with the superexchange interaction $J_\text{ex}$ and are much less coherent than in the case of the low-frequency excitations. 

\begin{figure}
  \includegraphics[width=1.0\linewidth]{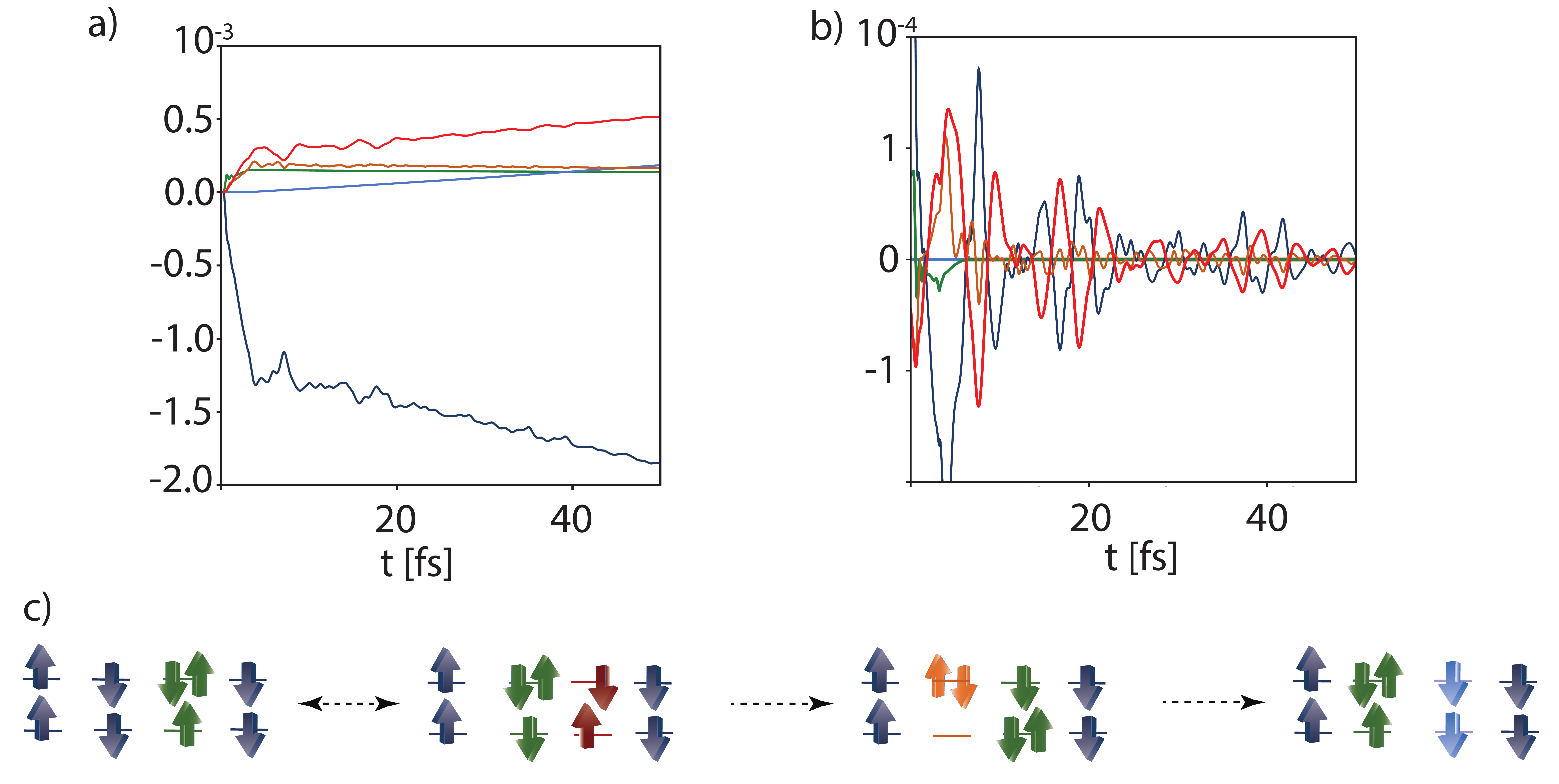}
  \caption{\textbf{a}, Time evolution
  of the occupation for the most relevant local many body states on the A sub-lattice, and
  \textbf{b}, the same data with subtracted background dynamics. The color lines match the
  graphical representation of the local many-body states in the main text. \textbf{c}, Schematic view of the decay channel for coherent oscillations. Transient coherent oscillations can decay via the creation of a high-energy Hund excitation S0$^{\flat}$ and the formation of a ferromagnetic domain wall.}
\label{SM:decay}
\end{figure}

The main difference with respect to the low-frequency excitation protocol is a rather fast damping of the oscillations which indicates a substantial role of incoherent processes which reduce their lifetime. This is even more evident in the analysis of the local many-body oscillations, see Fig.~\ref{SM:decay}(a) and (b), where transient oscillations between the ground state and the Hund excitation S0 are damped. In contrast to the low-frequency excitation, the probability for spin excitations S1$^{\flat}$ is steadily increasing, which marks the dynamical creation of ferromagnetic domain walls in the AFM background. In Fig.~\ref{SM:decay}(c), we sketch a high-order process that leads to the enhancement of the ferromagnetic domain walls and the decay of coherent oscillations. This theoretical result suggests that an analysis of the lifetime of the coherent oscillations with respect to the pump photon energy is an interesting experimental avenue to explore the coupling of the coherent and incoherent processes in systems with coupled Hund and antiferromagnetic physics.

\end{document}